\documentclass{article}

\usepackage{arxiv}

\usepackage[utf8]{inputenc} % allow utf-8 input
\usepackage[T1]{fontenc}    % use 8-bit T1 fonts
\usepackage{hyperref}       % hyperlinks
\usepackage{url}            % simple URL typesetting
\usepackage{booktabs}       % professional-quality tables
\usepackage{amsfonts}       % blackboard math symbols
\usepackage{nicefrac}       % compact symbols for 1/2, etc.
\usepackage{microtype}      % microtypography

\usepackage{cite}
\usepackage{color}
\usepackage{amsmath,amssymb,amsfonts}
\usepackage{graphicx}
\usepackage{textcomp}
\usepackage{algpseudocode}
\usepackage{algorithmicx,algorithm}
\usepackage{tabularx}
\usepackage{subfigure}
\usepackage{multirow}
\usepackage{amsmath,bm}
\usepackage[justification=centering]{caption}
\usepackage{booktabs}
\usepackage{amssymb}
\usepackage{grffile}

\newtheorem{theorem}{\hskip 1em  Theorem}

\newtheorem{definition}{\hskip 1em Definition}

\title{Local Causal Structure Learning and its Discovery Between Type 2 Diabetes and Bone Mineral Density}

\author{
    Wei Wang\\
    Division of Life Sciences and Medicine\\
    University of Science and Technology of China\\
    Hefei, Anhui 230001, P.R. China \\
    \texttt{hfww2001@fsyy.ustc.edu.cn} \\
     \And
    Gangqiang Hu\\
    School of Computer Science and Technology\\
    University of Science and Technology of China\\
    Hefei, Anhui 230001, P.R. China \\
    \texttt{stariate@ee.mount-sheikh.edu} \\
     \AND
    Bo Yuan \\
    Department of Computer Science\\
     Southern University of Science and Technology\\
     Shenzhen, P.R. China. \\
   \texttt{yuanb@sustc.edu.cn} \\
   \And
   Shandong Ye \\
   Division of Life Sciences and Medicine\\
   University of Science and Technology of China \\
   Hefei, Anhui 230001, P.R. China \\
   \texttt{ysd196406@163.com} \\
   \And
   Chao Chen \\
   Division of Life Sciences and Medicine\\
   University of Science and Technology of China \\
    Hefei, Anhui 230001, P.R. China \\
    \texttt{chengchao@medmail.com.cn} \\
     \And
   YaYun Cui \\
   Division of Life Sciences and Medicine\\
   University of Science and Technology of China \\
    Hefei, Anhui 230001, P.R. China \\
    \texttt{yayuncui@163.com} \\
     \And
   Xi Zhang \\
   Division of Life Sciences and Medicine\\
    University of Science and Technology of China \\
    Hefei, Anhui 230001, P.R. China \\
    \texttt{1518867079@qq.com} \\
     \And
   Liting Qian \\
   Division of Life Sciences and Medicine\\
   University of Science and Technology of China \\
   Hefei, Anhui 230001, P.R. China \\
    \texttt{money2004@sina.com} \\
}

\begin{document}
\maketitle

\begin{abstract}
Type 2 diabetes (T2DM), one of the most prevalent chronic diseases,
affects the glucose metabolism of the human body, which decreases the
quantity of life and brings a heavy burden on social medical care.
Patients with T2DM are more likely to suffer bone fragility fracture
as diabetes affects bone mineral density (BMD). However, the
discovery of the determinant factors of BMD in a medical way is
expensive and time-consuming. In this paper, we propose a novel
algorithm, \textbf{P}rior-\textbf{K}nowledge-driven local
\textbf{C}ausal structure \textbf{L}earning (PKCL), to discover the
underlying causal mechanism between BMD and its factors from the
clinical data. Since there exist limited data but redundant prior
knowledge for medicine, PKCL adequately utilize the prior knowledge
to mine the local causal structure for the target relationship.
Combining the medical prior knowledge with the discovered causal
relationships, PKCL can achieve more reliable results without
long-standing medical statistical experiments. Extensive experiments
are conducted on a newly provided clinical data set. The experimental
study of PKCL on the data is proved to highly corresponding with
existing medical knowledge, which demonstrates the superiority and
effectiveness of PKCL. To illustrate the importance of prior
knowledge, the result of the algorithm without prior knowledge is
also investigated.\end{abstract}

% keywords can be removed

\section{Introduction}

Diabetes mellitus is one of the most common chronic
diseases featured by high levels of blood glucose and type 2
diabetes mellitus (T2DM) is the most frequent subtype of diabetes
mellitus. T2DM and its complications cause a variety of health
problems and they bring heavy economic burdens to individuals
worldwide \cite{zhou2016worldwide}. Osteoporosis is a common
skeletal system disease characterized by decreased bone density and
normal bone microstructure deterioration predisposing to an
increased risk of bone fracture \cite{ghodsi2016mechanisms}.
Osteoporosis leads to a decrease in physical function and the
impairment of quality of life. Moreover, bone fracture due to
osteoporosis causes increased disability rate, mortality, and a great
economic burden on family and society \cite{cauley2017osteoporosis}.

Measurement of bone mineral density by dual X-ray
absorptiometry (DXA) is the most commonly used approach to diagnose
osteoporosis\cite{kanis2000update}. Decreased BMD reflects the
reduction in bone strength that is closely linked to increased bone
fracture risk. Osteoporosis-related bone fracture frequently occurs
in patients with
T2DM\cite{ghodsi2016mechanisms,goldshtein2018epidemiology}. Notably,
although patients with T2DM have higher risks of
osteoporosis-related bone fracture than those in non-diabetic
individuals, the BMD is not necessarily
identical\cite{deshields2018comparison,muschitz2019diagnosis}. As
suggested in a recent meta-analysis by Vestergaard, BMD even
increases in patients with T2DM compared with non-diabetic
individuals\cite{vestergaard2007discrepancies}.

Many factors affect BMD in diabetes conditions. The traditional
large longitudinal prospective studies are helpful to unravel
determinant factors of BMD in T2DM. However, these kinds of studies
are very expensive in terms of cost and time that they are difficult
to reach the conclusion within a short time. In addition, the
studies on the determinants of BMD in T2DM need to carry out
complicated data analyses and data processing due to the complexity
and complications of T2DM. Existing methods to find the relationship
between risk factors and BMD mostly rely on experts' knowledge and
artificial analysis of clinical data, which is time-consuming and
cost-effective. Furthermore, they cannot identify the underlying
causal mechanism between risk factors and BMD in T2DM.

To automatically identify the risk factors of BMD and discover the
underlying casual mechanism among them, intelligent algorithms should
be developed. Traditionally, Bayesian networks (BN) structure
learning algorithms can learn the casual mechanism from the data.
However, in the medical field, the number of clinical samples are
not enough for a BN structure learning algorithm to discover the
real underlying causal mechanism. Moreover, as BMD is affected by
numerous factors, traditional BN structure learning algorithms can
not be applied to such a large scale of factors. Considering that
lots of existing medical knowledge are not exploited, this paper
proposes a new BN structure learning algorithm (PKCL), which can
learn the underlying causal mechanism between BMD and it's factors,
meanwhile, incorporating rich existing prior knowledge. With the advantage of incorporating prior knowledge when learning the BN structure,
some of the parameters of the model are determined by the prior knowledge. 
Thus, PKCL can deal with the case of large number of factors. Benefiting
from prior knowledge, PKCL provides insight into complicated diseases
and offer useful information to clinical experiment. Our
contributions are summarized into the following three aspects:
\begin{enumerate}
    \item Aiming to the clinical data with scarce samples but abundant prior knowledge, a new framework is present to learn a more accurate model.
    \item  A structure learning algorithm, PKCL, is proposed to utilize the prior knowledge as well as the causal information to detect the causal relationships in clinical data.
    \item  We conclude the prior knowledge of experts about BMD and its risk factors. Conditioned on that, we discover the underlying causal mechanism between BMD and risk factors.
\end{enumerate}

\section{Related Work}

It is accepted that patients with T2DM have a higher risk of osteoporosis-related
bone fracture than those without diabetes\cite{janghorbani2007systematic,bonds2006risk,goldshtein2018epidemiology}. Measurement of BMD is used to diagnosing osteoporosis as the golden standard.
Nevertheless, whether the BMD decreases in T2DM is paradoxical according to current clinical studies.

A number of factors affect the BMD in diabetes conditions, such as sex, body mass index (BMI), insulin, and glucose.
The prevalence of higher BMD in T2DM is similar in men and women across racial and ethnic groups
including Mexican American, white, and black people\cite{ma2012association,strotmeyer2004diabetes,kao2003type}.
BMI is strongly associated with BMD in T2DM and might explain, in part, higher BMD in T2DM compared
with non-diabetic individuals\cite{shanbhogue2016type}. Insulin resistance and hyperinsulinemia,
which are characteristics of T2DM, have effects on bone metabolism. High levels of circulating
insulin may contribute to high BMD and there are evidences in preclinical models that altered insulin
levels and insulin resistance affect bone remodeling via direct effects on osteoblasts, osteoclasts,
and osteocytes, all of which express insulin receptors\cite{fulzele2010insulin}. Hyperglycemia is
associated with the accumulation of advanced glycation end-products (AGEs) in the bone matrix, and AGEs
inhibit bone formation, an effect mediated at least in part by increased osteocyte sclerostin
production\cite{tanaka2015effects,compston2018type}. Given the determinants of BMD is complicated,
the derivation of causality will contribute to elucidate the cause of bone mineral density in T2DM,
which is beneficial to prevent and treat osteoporosis-related bone fracture in T2DM.

However, the current work about selecting the most relative risk factors is rarely studied.
The existing approaches are mainly depended on the analysis and experience of the experts,
which are not cost-effective and time-efficient. In addition, they can't analyze the risk
factors of a complicated disease from a data aspect.

In recent studies, feature selection (FS) has been applied to
several tasks including classification, regression, and clustering.
A number of FS methods
\cite{jiang2019probabilistic,WU,jiang2019joint,he2015robust}, which
exploit different criteria to select the most informative features,
have been proposed in the literature. They can roughly be divided into
three classes: filter, wrapper, and embedding
methods\cite{liu2007computational}. However, these three classes can
not discover the underlying causal relationship between features and
targets. Moreover, their FS criteria lack a theoretical proof of the
optimality. The Markov Blanket (MB) algorithms are showed to have a superior performance
over the traditional FS algorithm, as the MB is proved to be the optimal
feature subset\cite{aliferis2010local,yu2019multi,yu2018mining}.  And MB algorithms can
discover the underlying causal mechanisms of the selected features
utilizing causal feature selection and causal discovery.

Generally, MB discovery can be grouped into two main types:
nontopology-based and topology-based. Nontopology-based MB algorithms
exploit independent tests between feature variables and target
variables to discover the MB heuristically. Koller-Sahami (KS)\cite{KS} 
first proposed an approximate algorithm to find the
MB, which minimizes the cross-entropy loss by pruning out some
redundant variables in a backward way. Due to the unsoundness of KS,
lots of nontopology-based algorithms are proposed to improve on it.
The Growth and Shrink algorithm (GS)\cite{GS} first tests and adds
variables, which are sorted by the mutual information with the
target variables, into the MB set in the growth stage. Then the
shrinking stage eliminates false-negative nodes from the previous MB
sets. Based on GS, The increment associated MB algorithm (IAMB)\cite{IAMB} 
improves the performance of GS by resorting the
variables each time the MB set changes. After that, numerous
variants of IAMB have been proposed including IAMBnPC, inter-IAMB,
and KIAMB\cite{PCMB}. However, with the size of variables growth,
the need for samples grows exponentially. If the sample data isn't
enough, the performance of IAMB and its variants will degrade.

As the limited data in real-world applications, topology-based
methods are proposed to solve the data efficiency while keeping a
reasonable time cost. Min-max MB (MMMB)\cite{MMMB} discovers the MB
set by finding the parent-and-children set first and then finding
the spouses, in which way the sample size only relies on the Directed Acyclic Graph (DAG)
structure rather than the size of variables. Although MMMB is later
proved to be unsound\cite{HITON-MB}, the two steps of discovering
the MB set are the foundation of the following methods. HITON-MB
\cite{HITON-MB} inherits the framework of MMMB and interweaving the
two steps to exclude the false positives from parents and children (PC) sets as early as
possible, which can decrease the number of independent tests (ITs)
needed later. However, Both MMMB and HITON-MB are unsound due to the
incorrectness of PC discovery. Parent children-based MB algorithm (PCMB)\cite{PCMB}, 
the first sound topology-based MB algorithm,
which utilizes a double check strategy to fix the errors in PC
discovery, is then introduced by Pena et. al. After that, Iterative
parent children-based MB (IPCMB) algorithm\cite{IPCMB} are proposed
based on PCMB and discover the PC set more efficiently. Recently,
Simultaneous MB algorithm (STMB)\cite{STMB} is developed to improve
the time efficiency of MB algorithms by utilizing the property of
coexisting between descendants and spouses.

Although MB algorithms can discover the underlying causal mechanism
between variables and targets, they can't recognize the direction of
the dependency. By BN structure learning, a DAG over all nodes can
be constructed using the local MB sets. One approach of learning BN
structure is constraint-based, which discover the arcs between each
node pairs by conditional independent test (CIs). However, the
number of CIs needed growth exponentially with the increase of the
nodes. Moreover, as each CIs is calculated based on the results of
another, it will lead to inevitable escalated errors. Another
approach to learning BN structure is score-based.

\section{Notation and Definition}

Let capital letters denote variables (such as $M,N$), lower-case letters (such as $m,n$) denote the value of random variables and capital bold italic (such as $\bm M,\bm N$) denote variable sets.

\begin{definition}[\textbf{Bayesian Network \cite{Notation-Bayesian}}]
Formally, a Bayesian Network is  a triplet $<\bm{G},\bm{P},\bm{U}>$, which denotes
a joint probability distribution $\bm{P}$ over a random variable set $\bm{U}$
and can be represented by a DAG where each node corresponds to a random variable.
\end{definition}

If there is an arc from $M$ to $N$, which means $MN \in G$, then $M$ is said to be
a $parent$ of $N$ and $N$ is a children of $M$. In addition, if $M$ is a $parent$
or $children$ of $N$,they are said to be $neighbors$. Node $M$ and $N$ are said to
be $spouses$ of each other if they have a common child and there is no arc between
$M$ and $N$.If there is a directed path from $M$ to $N$ in G, then $N$ is a descendant
of $M$. And the descendants and the parents of $X$ is represented as $Des(X)$
and $Pa(X)$. Further,  we use $H^G(X)$ and $S^G(X)$ to denote the neighbors and the
spouse of node $X$ in $G$.

\begin{definition}[\textbf{Markov Condition\cite{Notation-Bayesian}}]
Every node in the BN is independent of its nondescendant nodes, given its parents.
Thus, if a BN $<\bm{G},\bm{P},\bm{U}>$, according to the definition of
Markov Condition, the joint probability $\bm{P}$ can be decomposed into the
product of a series conditional probabilities:
$$P(U)=\prod_{X \in U}P(X|Pa(X))$$

\end{definition}

\begin{definition}[\textbf{V-Structure\cite{Notation-Bayesian}}]
Three nodes $M$, $X$, and $N$ are said to be a V-structure if there are two
arcs from $M$, $N$ to $N$ and $M$ is not adjacent to $X$.
\end{definition}

$X$ is said to be a collider if $X$ has two incoming arcs from $M$ and $N$,
no matter $M$ and $N$ are adjacent or not. On the condition that  $M$ and $N$
are adjacent, we say $Y$ is an unshielded collider for the path from $M$ to $N$.

\begin{definition}[\textbf{Blocked Path\cite{Notation-Bayesian}}]
Any path from node in $\bm M$ to node in $\bm N$ is said to be blocked by a
variable set $\bm{\Phi}$ iff: 1) ${\bm\Phi}$ comprises a head-to-tail
($M \rightarrow X \rightarrow N$) or tail-to-tail ($M \leftarrow X \rightarrow N$)
chain, and $M \in \bm N$. 2) ${\bm\Phi}$ comprises a head-to-head
($M \rightarrow X \leftarrow N$) chain, where $X \notin \bm Z$ and any node in
$Des(X) \notin \bm Z$.
\end{definition}

\begin{definition}[\textbf{d-Separation\cite{Notation-Bayesian}}]
If all paths from $\bm M$ to $\bm N$ is blocked by ${\bm\Phi}$, then ${\bm\Phi}$
is said to d-sperate $\bm M$ and $\bm N$, denoted as $Dsep(M,N|\bm Z)$
\end{definition}

\begin{definition}[\textbf{Faithfulness Condition\cite{Notation-Bayesian}}]
Given a BN $<\bm{G},\bm{P},\bm{U}>$, $\bm P$ and $\bm G$ are faithful to
each other iff: all and only the condition probabilities true in $\bm P$
are entailed by $\bm G$. Formally, for any $M,N$ in $\bm U$ and
$\bm Z \subseteq U-\{M,N\}$, $M\perp N | \bm Z $ in $\bm P$
iff $Dsep(M,N|Z)$ in $\bm G$
\end{definition}

\begin{definition}[\textbf{Markov Blanket\cite{Notation-Bayesian}}]
Formally, given the MB of a target node $T$, denoted as $MB(T)$, $T$ is
independent of $\bm U \setminus MB(T)$.
\end{definition}

\begin{definition}[\textbf{PCMasking\cite{CCMB}}]
Let $PC(X)$ denotes the PC set of variable X. $PC_{1}$ and $PC_{2}$ denote  two
subsets of $PC(X)$ and $PC_{1} \cap PC_{2} = \varnothing$. $PC_{1}$ and $PC_{2}$
are PCMaksing for variable X if
$$X \perp PC_{1}|PC_{2}, X \perp PC_{2}|PC_{1}$$
$PC(X)$ and $PC_{1}$ are called MaskingPCs.
\end{definition}

\begin{theorem}[\textbf{MB Uniqueness\cite{Notation-Bayesian}}]
Given a BN$<\bm{G},\bm{P},\bm{U}>$, if $\bm P$ and $\bm G$ are faithful to
each other, then $ MB(T),T \in \bm U$, is unique and is the node set of neighbors
$H(T)$ and spouses $S(T)$.In addition, $H(T)$ is also unique.
\end{theorem}

\begin{table*}[htbp]
\centering
\caption{STATISTICAL INFORMATION OF THE CLINICAL DATA SET. SIX KINDS OF BMDS ARE NAME OF BMD1 TO BMD6}
 \resizebox{\textwidth}{!}{
 \begin{tabular}{cccccccc}
  \toprule
  No.& Features & Type & Description & No.& Features & Type & Description\\
  \midrule
 1& Duration & Numeric & Duration of disease & 18 & AST & Numeric & Aspartate aminotransferase\\
 2& Sex& Boolean & & 19 & GGT & Numeric & Gamma glutamyltransferase\\
 3& Age& Numeric & & 20 & 20-OH-VD & Numeric & 25-hydroxyvitamin vitaminD\\
 4& Height& Numeric & & 21 & UALB/Ucr & Numeric & Urine albumin creatinine ratio \\
 5& Weight& Numeric & & 22 & BMD1 & Numeric & Lumbar spines (L1¨CL4)\\
 6& BMI& Numeric & BMI=weight(kg)/height(m$)^2$ & 23 & BMD2& Numeric &Distal radius \\
 7& FPG& Numeric &Fasting plasma glucose & 24 & BMD3  & Numeric & Femoral neck\\
 8& HbAlc& Numeric & Glycated hemoglobin & 25 & BMD4  & Numeric & Wards triangle\\
 9& Cr& Numeric & Serum creatinine& 26 & BMD5  & Numeric & Greater trochanter\\
 10& UA& Numeric &Serum uric acid & 27 & BMD6  & Numeric & Total hip\\
 11& Ca& Numeric &Calcium& 28 & OC & Numeric & Osteocalcin\\
 12& P& Numeric &Phosphorus & 29 & CTX & Numeric &  C-terminal telopeptide of type I collagen\\
 13& ALP& Numeric & Alkaline phosphatase& 30 & PINP & Numeric & N-terminal propeptide of type 1 procollagen\\
 14& TG& Numeric & Triglyceride& 31 & SBP & Numeric & Systolic blood pressure\\
 15& TC& Numeric & Total cholesterol& 32 & DBP & Numeric & Diastolic blood pressure\\
 16& Alb& Numeric & Albumin & 33 & LDL-C & Numeric & low-density lipoprotein cholesterol\\
 17& ALT& Numeric & Alanine aminotransferase& 34 & HDL-C & Numeric & High-density lipoprotein cholesterol\\

  \bottomrule
  \label{table1}
  \end{tabular}
  }
\end{table*}

\section{Methods}

In this section, we propose a BN structure learning algorithm driven by prior knowledge.
Section V-A demonstrates the structure of PKCL and Section V-B, Section V-C demonstrate two-stage of it.
\subsection{Overview}
%motivation
In real-world applications, the number of samples is limited while the number of features
is numerous. If directly develop a structure learning (SL) algorithm in the limited data set, the output DAG
can hardly reflect the real underlying casual mechanism among variables. Meanwhile, the
existing SL algorithms ignore the significance of experts' prior knowledge, which leads
to the poor performance of the algorithms. Motivated to incorporate the  SL algorithm with the
experts' prior knowledge, we propose an SL algorithm, which learns the BN structure and adds
the prior knowledge simultaneously to build a global structure.

PKCL algorithm works in two phases: the local stage and the global stage. The pseudo-code
of  PKCL algorithm is denoted as Algorithm 1. In the local stage (lines 1-4 of Algorithm 1),
PKCL first discovers the neighbors of the target variables and then detects the MaskingPCs to eliminate the effect of them. After that, it finds the spouse of target variables utilizing
the neighbors set. Thus, the skeleton of BN is constructed and the detail of this stage is
discussed in Section V-B.

In the global stage (lines 6-9 of Algorithm 1), PKCL leverages the MB sets learned in the
local stage to learn the global BN structure, in which prior knowledge is incorporated to
guide the global learning phase. Specifically, it learns the casual direction between feature
variables and target variables by combining the constraint-based method and score-based method.
What's more, in the learning phase, it automatically adds casual direction according to the
prior knowledge. The detail of this stage is discussed in Section V-C.

\begin{algorithm}[t]
\caption{PKCL}
\hspace*{0.02in} {\bf Input:}
    Data D on node set $\bm N$ , Target node set  $\bm T$, Prior rule set $\bm R$\\
\hspace*{0.02in} {\bf Output:}
    Directed acyclic graph $\bm G$
\begin{algorithmic}[1]
\State \{Local stage\}
\ForAll{ $T \in \bm T$ }
    \State $\bm{MB}(T)\gets$ \Call{CCMB}{$D,T$}
\EndFor

\State \{Global stage\}
\State $\bm G,flag \gets $\Call{FindCollider}{$D,\bm{MB},\bm R$}
\If{not $flag$}
    \State $\bm G \gets $\Call{ScoreSearch}{$D,\bm{MB},\bm R$}
\EndIf

\State \Return $\bm G$
\end{algorithmic}
\end{algorithm}

\subsection{The Local Stage}
In this stage, we present a cross-check and complement MB discovery (CCMB)\cite{CCMB}.
CCMB is a topology-based MB discovery algorithm,
different from other previews algorithm, it discovers the MB set of a node while repair
the incorrect conditional independent (CI) tests via eliminating the PCMasking phenomenon.

The pseudo-code of CCMB is represented in algorithm 2. Specifically, it works in the
following three steps.

Step1 (algorithm 3): Discover the neighbors of node $T$. The pseudo-code of this step is
represented in algorithm 3. For each target node $T$ in $\bm{T}$, the algorithm works in
three phases: First, find the potential neighbors set $\bm H(T)$ of $\bm T$, then score
and rank the potential neighbors to choose the best one from $H(T)$, and finally prune out
the false variables.

Step2 (lines 4-11 of algorithm 2): Prune out MaskingPCs of node $T$. Compared to other
MB algorithms, this step is the key point that makes CCMB outperform them. CCMB
exploits a cross-check method(lines 4-8) to discover the MaskingPCs and appends them
into PCMasking table $\bm{PCM}$ in the format of $[T, X]$, where $T$ denotes the target
variable and $X$ denotes the cross-checked variable. Specifically, if $T$ is the neighbor
of $X$ while $X$ is not the neighbor of $T$, the cross-check method will take $X$ and $T$
as MaskingPCs because of the asymmetry between them.

Step3 (algorithm 4): Discover the spouses of node $T$. The pseudo-code of this step
is represented in algorithm 4. If $Y$ is the neighbors of $T$ and $X$ is the neighbor
of $Y$, then find a node subset $\bm Z$ conditioned on which $T$ is independent of $X$
.

\begin{algorithm}[t]
\caption{CCMB(D,T)}
\hspace*{0.02in} {\bf Input:}
    Data D on node set $\bm N$ , Target node $T$\\
\hspace*{0.02in} {\bf Output:}
    The Markov boundary of $T$
\begin{algorithmic}[1]
\State \{Step 1: Find the Neighbors\}
\State $\bm H(T)\gets$ \Call{FindNeighbors}{$D,T$}
\State \{Step 2: Eliminate PCMasking phenomenon\}
\ForAll{$X \in \bm{N}-\{T\}$}
    \If{$T \in$ \Call{FindNeighbors}{$D,X$} and $X \notin \bm{H}(T)$}
        \State $\bm{PCM(V)} \gets \bm{PCM}(V) \cup \{[T,X]\}$
    \EndIf
\EndFor
\ForAll{$[T,X] \in \bm{PCM}(T)$}
    \State $\bm H(T)=\bm H(T) \cup \{X\}$
\EndFor
\State \{Step 3: Find the spouse set\}
\State $\bm S(T) \gets$ \Call{FindSpouse}{$D,T,\bm H(T)$}
\State \Return $\bm{MB}(T) \gets \bm{H}(T) \cup \bm{S}(T)$
\end{algorithmic}
\end{algorithm}

\begin{algorithm}[t]
\caption{FindNeighbors(D,$\bm T$):}
\hspace*{0.02in} {\bf Input:}
    Data D on node set N, Target node $\bm T$\\
\hspace*{0.02in} {\bf Output:}
    The PC subset
\begin{algorithmic}[1]
    \State \{ Step 1: Find the potential neighbors\}
    \ForAll{$T \in \bm{T}$}
        \State $\bm{H}(T) \gets \varnothing$, $\bm{PH}(T) \gets \bm N \setminus T$
        \While{$\bm{PH}(T) \neq \varnothing$}
            \ForAll{$X \in \bm{PH}(T)$}
                \State $\bm{Sep}[X]= arg \min_{\bm{Z}\subseteq \bm{H}}{dep(T,X|\bm{Z})}$
                \If{$T\perp X|\bm{Sep}[X]$}
                    \State $\bm{PH}(T) \gets \bm{PH}(T)-\{X\}$
                \EndIf
            \EndFor
            \ForAll{$X,Y \in \bm{PH}(T)$}
                \If{$X \not\perp Y$ and $T \perp X|Y$}
                    \State $\bm{PH}(T) \gets \bm{PH}(T)-\{X\}$
                \EndIf
            \EndFor
            \ForAll{$X \in \bm{PH}(T)$}
                \State $Score[X]=dep(T,X|\bm{Sep}[X])$
            \EndFor
            \State $Y=arg\max_{\bm{X}\in\bm{CanPC}}{Score}[X]$
            \State $\bm{H}(T) \gets\bm{H}(T) \cup \{Y\}$
            \State $\bm{PH}(T) \gets \bm{PH}(T) -\{Y\}$
            \ForAll{$X \in \bm{H}(T)$}
                \If{$\exists \bm{Z} \subseteq \bm{H}(T)-\{X\} s.t. T \perp X |\bm{Z} $}
                    \State $\bm H(T)\gets \bm H(T) -\{X\}$
                \EndIf
            \EndFor

        \EndWhile
    \EndFor
\State \Return The Neighbors set of nodes in $\bm T$ $\bm H$
\end{algorithmic}
\end{algorithm}

\begin{algorithm}[t]
\caption{FindSpouse(D,T,$\bm{H}$):}
\hspace*{0.02in} {\bf Input:}
    Data D on node set N, Target node T, MB sets\\
\hspace*{0.02in} {\bf Output:}
    The spouse set of $T$
\begin{algorithmic}[1]
\ForAll{$Y \in \bm H(T)$}
    \ForAll{$X \in \bm H(Y)$}
        \If{$X \notin \bm H(T)$}
            \State find $ \bm Z s.t. T \perp X |Z$ and $T,X \notin \bm Z$
            \If{$T\not \perp X | \bm{Z} \cup \{Y\}$}
                \State $\bm{S}(T) \gets \bm{S}(T) \cup \{X\} $
            \EndIf
        \EndIf
    \EndFor
\EndFor
\State \Return $\bm{S}$(T)
\end{algorithmic}
\end{algorithm}

\begin{algorithm}[t]
\caption{FindCollider($D, \bm{MB},\bm{R}$):}
\hspace*{0.02in} {\bf Input:}
    Data D on node set $\bm N$, $\bm MB$ sets, Prior rules $\bm R$\\
\hspace*{0.02in} {\bf Output:}
    Directed acyclic graph G
\begin{algorithmic}[1]
\State $\bm G \gets \varnothing $
\ForAll{$X \in \bm{N}$}
    \ForAll{$Y \in \bm{MB}(X)$}
        \ForAll{$Z$ is the common child of $X$ and $Y$}
            \If{possible without introducing cycles and satisfies prior knowledge R}
                \State add XY and YZ to $\bm G$
            \EndIf
        \EndFor
    \EndFor
\EndFor
\State \Return $\bm{G}$
\end{algorithmic}
\end{algorithm}

\subsection{The Global Stage}
After the MB sets discovered, local information can be integrated to get the structure of DAG.
Traditionally, the next step is to determine the direction of the edge, and thus, the underlying
causal mechanism is learned. However, this way of learning the DAG is totally depended on the clinical
data, which means a lot of knowledge in the medical field are ignored. Thus, some of the causal relationships that
only learned from clinical data in conflict with medical knowledge and some causal relationship that is already
proved in medical literature can not be learned.

PKCL learns the structure of a DAG between nodes via leveraging the MB sets discovered in the local state.
Different from other structure learning method, the learning process of PKCL is routed by the prior
knowledge of experts, which means the PKCL works in a more data-efficient way while maintaining superior performance. Specifically, PKCL first discovers the colliders to constructed the overall DAG. if there is no collider discovered, we use a heuristic method with the constraints of MB sets and prior rules to construct the DAG of the underlying BN. Here, the heuristic algorithm we used is the steepest ascent hill-climbing with a TABU\cite{tsamardinos2006max} list of the last 100 structures and a stopping criterion of 15 steps without improvement in the maximum score.

\section{Experiments}

\subsection{Data Collection and discretization}

In our study, the clinical data of patients with T2DM are collected in the Department of
the First Affiliated Hospital of the University of Science and Technology of China.
As some patient samples contain several missing values and abnormal values, the data set
is cleaned and completed during the preprocessing procedure. After that, a clinical data
set of PKCL patient sample are collected, in which each sample has 34 features including
anthropometric indexes of patients, biochemical indexes, lipid profile, and vitamin D levels
from the assay of patients¡¯ blood samples. And the 34 features are labeled from 1 to 34.
The description of them is shown in Table \ref{table1}. The prior knowledge used in the experiment
is that feature 3, 5, 6, 7, and 8 are the causes of six BMDs. Before experiments, the data should be discretized. Here we use a packed toolkit Causal Explorer\cite{statnikov2010causal}. The detailed description is added in the appendix.

\subsection{Quality analyzation of selected features}
The overall experiment comprises two-stage. To demonstrate the superiority of PKCL, each stage of
PKCL is analyzed. In the local stage, four traditional feature selection algorithms and four MB algorithms
are also applied to the clinical data set. To evaluate the quality of selected features, five classifiers are
learned based on the selected features of nine algorithms and the prediction accuracy is computed.
In the global stage, to demonstrate that the casual relationship learned after incorporated prior knowledge
is more reasonable than the causal relationship learned without prior knowledge, not only the DAG with prior
knowledge is learned, but also the DAG without prior knowledge are learned.

At first, we randomly select 400 samples from the dataset to implement the CCMB algorithm
and other four MB algorithms and four traditional feature selection algorithms, namely IAMB \cite{IAMB},
PCMB \cite{PCMB}, MBOR, STMB \cite{STMB}, mRMR, Fisher, FCBF, and RFS. Then, in order to demonstrate the superiority of PKCL, five classifiers, i.e., Support Vector Machine (SVM), k-Nearest Neighbors(kNN), AdaBoost,
Random Forest (RF), and Naive Bayes (NB) are trained with their selected features. In addition,
the classifiers are also trained with the original features to be considered as a baseline,
which can demonstrate that the feature selection algorithms can improve the prediction accuracy of
classifiers by extracting the informative features. The k is set to 10 in kNN classifier. Lastly,
the rest 100 samples are used as testing data to evaluate the quality of the selected features.

The experimental results on six BMD are listed in Table \ref{table2}. As Table \ref{table2} shows,  when the label is BMD1, BMD2 or BMD5, the five classifiers using the features selected by
PKCL achieves the best prediction accuracy.  When the label is BMD3, BMD4 or BMD6, SVM, KNN, Adaboost, and Random Forest also achieves the best prediction accuracy using the feature selected by PKCL,
although Naive Bayes achieves the best prediction using the feature selected by mRMR, the result is
still competitive when using the feature selected by PKCL. Specifically, the five classifiers achieve
0.9-19.2\% improvement of prediction accuracy in comparison to the result of using all features, which
brings a significant improvement. In addition, the selected features are the input of the global stage of
PKCL, if the selected features are more informative, the underlying causal mechanism will be more reasonable.

\begin{table*}[htbp]
\centering
\caption{The prediction accuracy (in \%) on the test samples. The best result of each classifier is illustrated in bold.
All denotes no feature selected algorithm is applied.
}
 \resizebox{\textwidth}{!}{ 
 \begin{tabular}{cccccccccccc}
  \hline
    Lable& Classify &IAMB&PCMB&MBOR&STMB&mRMR&Fisher&FCBF&RFS&ALL&Ours \\
  \hline
    \multirow{5}{*}{BMD1}  &SVM&55.3&65.283&60.427&63.75&72.354&69.423&67.941&68.014&64.57&\textbf{76.219} \\
                        &KNN&58.393&69.6&60.646&67.087&74.127&70.086&72.467&69.929&66.522&\textbf{80.175} \\
                        &Adaboost&55.915&67.009&60.444&64.187&68.64&66.923&68.02&67.546&63.123&\textbf{72.875} \\
                        &Random Forest&59.013&69.038&60.921&68.167&75.773&74.146&72.909&70.628&63.151&\textbf{78.903}\\
                        &Naive Bayes&59.475&68.923&60.581&67.396&74.339&68.131&72.458&71.278&65.027&\textbf{75.242}  \\
    \hline
    \multirow{5}{*}{BMD2}
                        &SVM&53.544&63.951&55.388&62.936&69.488&65.762&63.926&63.67&61.005&\textbf{73.201} \\
                        &KNN&56.633&64.847&58.898&64.916&69.927&65.494&71.202&67.628&62.25&\textbf{77.015} \\
                        &Adaboost&53.108&64.446&56.529&62.39&64.747&64.319&63.781&63.329&62.322&\textbf{71.599}  \\
                        &Random Forest&54.081&65.209&56.983&63.644&73.259&71.095&71.798&67.62&61.656&\textbf{77.618}  \\
                        &Naive Bayes&56.168&65.689&58.777&64.827&72.044&65.393&68.061&67.106&62.085&\textbf{73.055}\\
    \hline

    \multirow{5}{*}{BMD3}  &SVM&54.093&63.519&57.711&62.894&69.368&67.011&65.427&65.038&62.501&\textbf{74.378}\\
                        &KNN&56.891&68.333&59.328&65.395&72.031&68.046&71.813&67.889&63.685&\textbf{78.917} \\
                        &Adaboost&53.4&63.928&57.85&63.386&66.364&65.656&65.904&65.571&61.473&\textbf{71.084} \\
                        &Random Forest&57.376&67.778&60.181&65.916&72.127&71.206&72.25&69.043&61.777&\textbf{76.949}\\
                        &Naive Bayes&56.704&66.208&59.913&65.701&\textbf{73.765}&65.993&69.108&68.93&61.975&72.93  \\
    \hline

    \multirow{5}{*}{BMD4}  &SVM&53.253&61.409&54.578&59.979&70.107&67.461&61.401&65.376&62.968&\textbf{70.197} \\
                        &KNN&53.747&67.956&57.149&63.724&73.254&69.152&67.877&64.256&60.841&\textbf{78.813} \\
                        &Adaboost&54.54&62.053&55.864&61.419&63.851&62.71&65.129&62.369&59.105&\textbf{67.217} \\
                        &Random Forest&56.345&66.37&60.107&61.976&73.471&70.079&67.521&66.696&59.252&\textbf{78.304} \\
                        &Naive Bayes&57.193&64.551&55.76&64.231&\textbf{71.444}&64.54&67.782&66.282&60.751&69.344 \\
    \hline

    \multirow{5}{*}{BMD5}  &SVM&50.583&61.595&56.123&59.385&66.177&64.351&61.391&64.528&59.907&\textbf{73.295}  \\
                        &KNN&52.876&68.313&59.261&63.652&72.242&67.172&71.41&65.663&63.704&\textbf{74.675} \\
                        &Adaboost&51.036&65.207&57.615&59.669&67.302&62.788&63.952&61.715&61.853&\textbf{70.117}  \\
                        &Random Forest&53.231&64.202&59.92&65.132&73.453&69.624&68.88&65.835&60.399&\textbf{75.876}\\
                        &Naive Bayes&53.374&65.446&59.815&64.959&72.456&63.077&68.431&66.953&62.478&\textbf{73.045} \\
    \hline

    \multirow{5}{*}{BMD6}  &SVM&53.711&64.191&57.831&62.89&69.616&65.743&64.496&66.123&62.75&\textbf{74.558} \\
                        &KNN&56.755&66.571&59.665&65.56&72.618&68.729&71.871&68.614&63.417&\textbf{78.743} \\
                        &Adaboost&54.615&64.692&57.83&62.673&65.816&65.955&66.14&65.566&62.76&\textbf{71.893} \\
                        &Random Forest&56.593&66.914&58.654&65.191&73.266&71.568&71.959&68.351&61.332&\textbf{76.743} \\
                        &Naive Bayes&57.065&66.75&59.95&65.592&\textbf{73.39}&65.815&70.713&70.173&62.404&72.326 \\
    \hline

  \label{table2}
 \end{tabular}
 }

\end{table*}

\subsection{Learning the DAG with prior knowledge}

\begin{figure*}[htbp]
\centering

\subfigure[BMD1]{
\begin{minipage}[t]{0.5\linewidth}
\centering
\includegraphics[width=8cm]{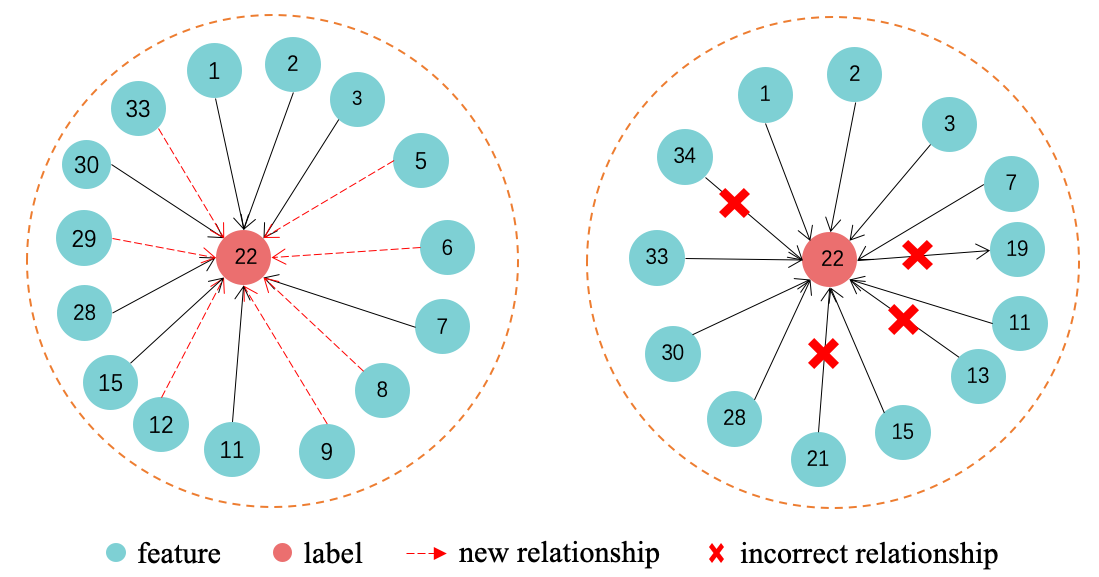}
%\caption{fig1}
\end{minipage}%
}%
\subfigure[BMD2]{
\begin{minipage}[t]{0.5\linewidth}
\centering
\includegraphics[width=8cm]{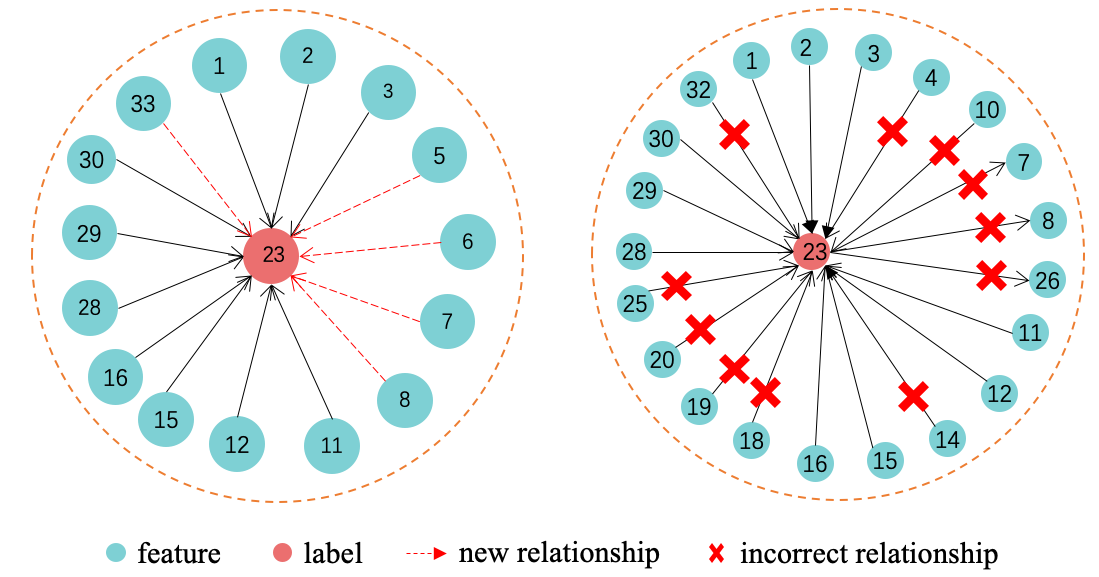}
%\caption{fig2}
\end{minipage}%
}%

\subfigure[BMD3]{
\begin{minipage}[t]{0.5\linewidth}
\centering
\includegraphics[width=8cm]{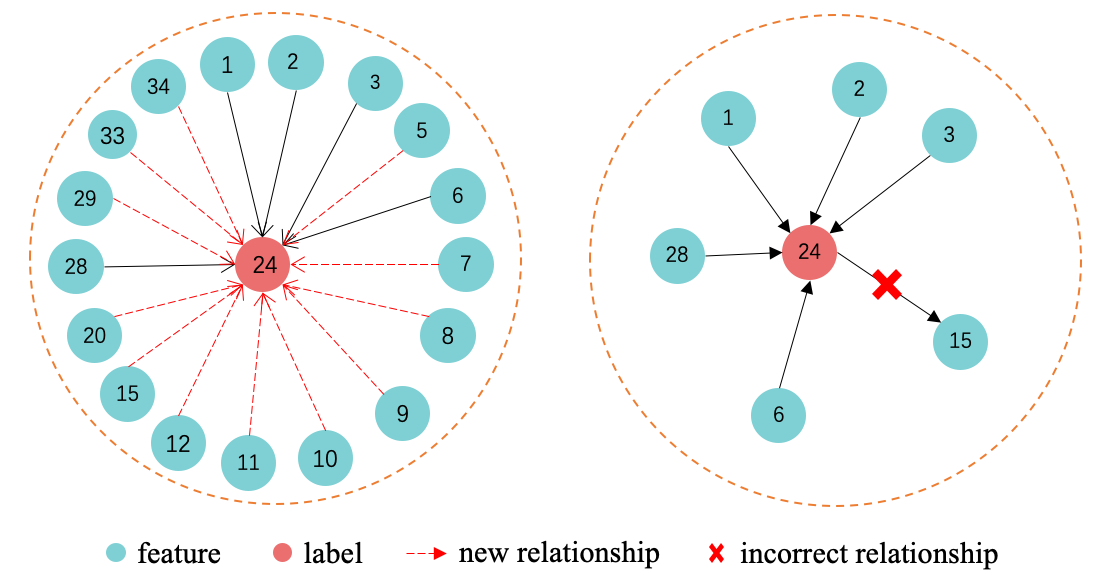}
%\caption{fig2}
\end{minipage}
}%
\subfigure[BMD4]{
\begin{minipage}[t]{0.5\linewidth}
\centering
\includegraphics[width=8cm]{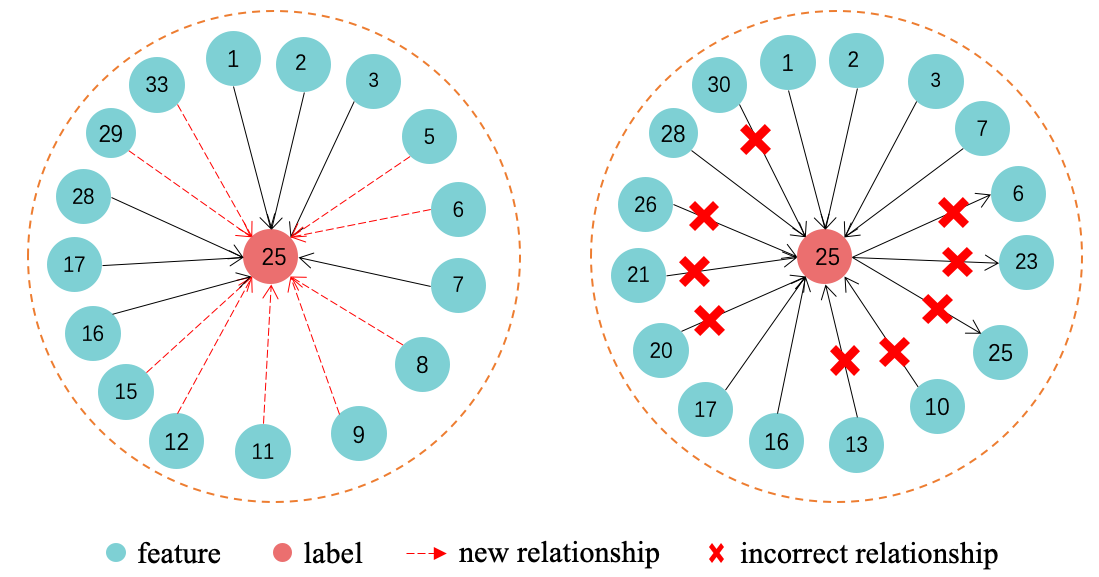}
%\caption{fig2}
\end{minipage}
}%

\subfigure[BMD5]{
\begin{minipage}[t]{0.5\linewidth}
\centering
\includegraphics[width=8cm]{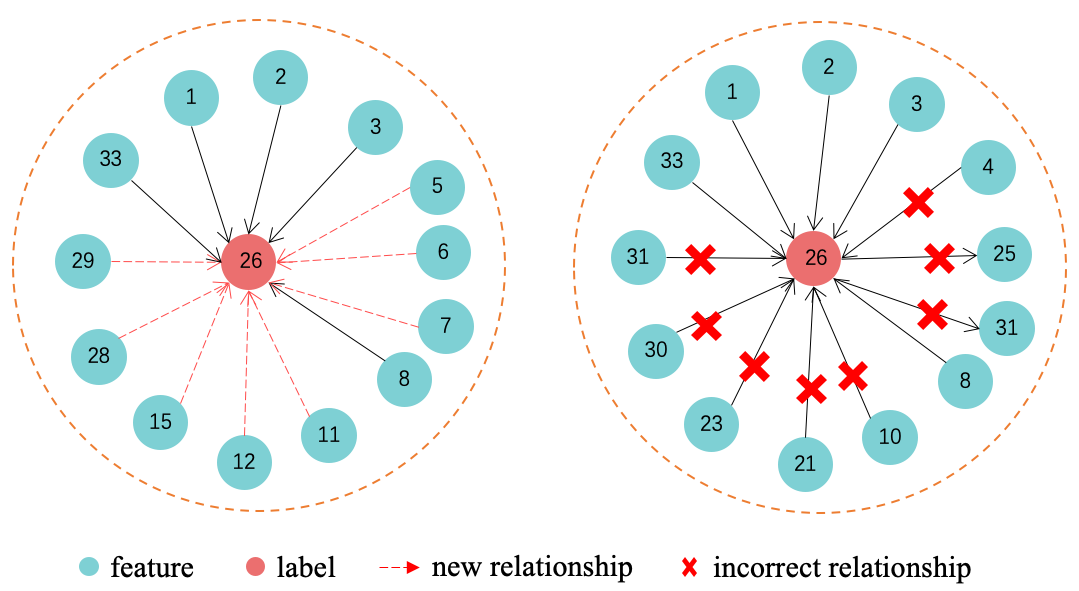}
%\caption{fig2}
\end{minipage}
}%
\subfigure[BMD6]{
\begin{minipage}[t]{0.5\linewidth}
\centering
\includegraphics[width=8cm]{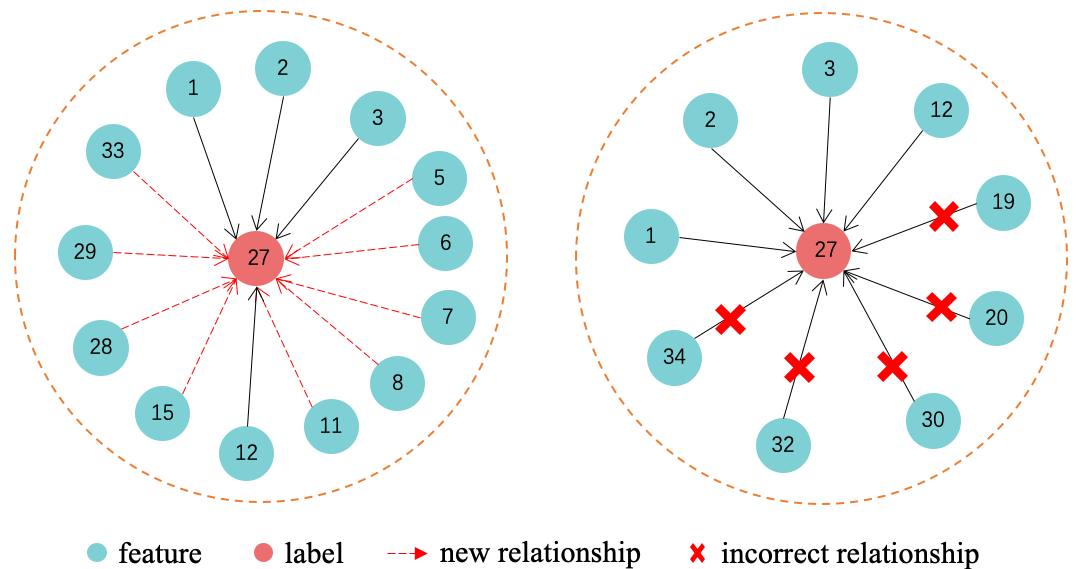}
%\caption{fig2}
\end{minipage}
}%

\centering
\caption{The causal relationships of six BMDs. Each sub-figure illustrates the local causal relationship of one
BMD. The left of each sub-figure is the local relationships that incorporating prior knowledge and the right is local relationships that not incorporating prior knowledge. The red circle in the central presents BMD and cyan circle in the edge presents feature.
}
\label{fig1}

\end{figure*}

To illustrate the significance of prior knowledge, the DAG that not incorporating prior
knowledge is also learned. The overall DAG learned with prior knowledge and the overall DAG
learned without prior knowledge are presented in the appendix. Here only six BMDs concerned
are analyzed. Figure \ref{fig1} is the local casual relationship of six BMDs and features.
Each sub-figure presents the local casual relationships of one BMD, which contains both the local casual
relationships incorporating prior knowledge and not incorporating prior knowledge.

In order to have an insight into the differences between the DAG that incorporating prior knowledge and the DAG that not incorporating prior knowledge, we first analyze the DAG that incorporating prior knowledge and then analyze the DAG not incorporating prior knowledge, finally, the superiority of the former and the inaccuracy of the latter are analyzed in detail.

The DAG incorporating prior knowledge is analyzed as follows. As Figure \ref {fig1} shows, all BMDs have no effect on any feature, which means BMDs are the comprehensive effects of some risk factors and BMDs don't have effects on any risk factors. In addition, feature 1, 2, 3, 5, 6, 7, 8, 11, 12, 15, 28, 29, 33 are the common causes of all six BMDs, which means these features have an underlying effect on the decrease of mine density. Feature 9,30 are the causes of BMD1. Feature 16, 30 are the causes of BMD2. Feature 9, 10, 20, 34 are the causes of BMD3. Feature 16, 17 are the causes of BMD4. The effect of BMD1 is feature 22. The effects of BMD2 are feature 7, 8, 26. The effect of BMD3 is feature 15. The effects of BMD4 are feature 6, 23, 25. The effects of BMD5 are feature 25, 32. BMD6 has no effect on any feature.

The DAG not incorporating prior knowledge is analyzed as follows. Feature 1, 2, 3  are the common causes of all six BMDs. Feature 11, 13, 15, 21, 28, 30, 33, 34 are the causes of BMD1. Feature 4, 10, 11, 12, 14, 15, 16, 18, 19, 20, 25, 28, 29, 30, 32 are the causes of BMD2. Feature 6, 28 are the causes of BMD3. Feature 7, 10, 13, 16, 17, 20, 21, 26, 28, 30 are the causes of BMD4. Feature 4, 8, 10, 21, 23, 30, 31, 33 are the causes of BMD5. Features 12, 19, 20, 30, 32, 34 are the causes of BMD6.

The difference between the DAG incorporating prior knowledge and not incorporating prior knowledge is analyzed as follows.  For BMD1, the arcs from feature 13, 21, 34 to BMD1 and the arc BMD1 from 19 are removed while there are new arcs added from features 5, 6, 8, 9, 12, 29, 33 to BMD1. The arc from feature BMD3 to feature 15 is removed while there are new arcs added from features 5, 7, 8, 9, 10, 11, 12, 15, 20, 29, 33, 34 to BMD3. The arcs from feature 7, 10, 13, 20, 21, 26, 30 from BMD4 and arcs from BMD4 to feature 6, 23, 25 are removed while there are new arcs added from feature 5, 6, 7, 8, 9, 11, 12, 15, 29, 33 to BMD4. The arcs from features 4, 8, 10, 21, 23, 30, 31 to BMD5 and arcs from BMD5 to 25, 31 are removed while there are new arcs added from feature 5, 6, 7, 11, 12, 15, 28, 29 to BMD5. The arcs from feature 12, 19, 20, 30, 32, 34 to BMD6 are removed while there are new arcs from feature 5, 6, 7, 8, 11, 12, 15, 28, 29, 33 to BMD6.

\subsection{Discussion}

As we analyzed above, PCKL can discover the underlying causal mechanism between BMDs and their related
risk factors. Here some casual relationships that have already been discovered in the clinical field will be
discussed, which can demonstrate the superiority of PKCL. In addition, the new casual mechanism found by
PCKL will provide an insight into the relationship between BMDs and their factors, which may contribute to the prevention and treatment of diabetes-related osteoporosis.

A report shows that 1 out of 3 women and 1 out of 5 men over 50 years old will experience
an osteoporotic fracture at some point in their life\cite{paschou2017type}. Patients with T2DM,
one of the most common chronic diseases, suffer from an increased osteoporotic-related bone fracture risk,
which places a heavy burden on individuals. BMD is the golden standard for diagnosing
osteoporosis. However, the causal chain involved in BMD and T2DM is not clear.

In elderly diabetic individuals, AGEs may inhibit the phenotypic expression of osteoblast
and promote osteoblast apophasis, thereby contributing to the deficiency in the bone
formation\cite{katayama1996role,alikhani2007advanced}. AGEs also increases osteoclast-induced
bone resorption. The study by Zhou et al. has indicated that increasing age is a more important
risk factor for bone mineral loss in patients with T2DM than diabetes duration\cite{zhou2010prevalence}.
The report by Wang et al. has indicated that the adverse changes in the collagen network occur with
aging and such changes may lead to the decreased toughness of the bone\cite{wang2002age}.
Moreover, the porosity of the bone significantly increases with aging and correlates to bone
strength and stiffness\cite{wang2002age}. Therefore, BMD negatively correlates to aging.
After bone mass reaches a peak in the third or fourth decade of life, vertebral bone mass
and density decrease with aging for both females and males\cite{ebbesen1999age}. Moreover, AGEs
accumulation occurs in the bone with aging, increasing by 4 to10 fold at the age, of 50 years
old\cite{illien2018dietary}. As discussed below, increased levels of AGEs in bone tissues have
been shown to be associated with diminished bone mechanical function and reduced cortical and
trabecular bone strength. Additionally, age-related bone loss is associated with abnormalities in
vitamin D status. Reduced serum levels of active vitamin D metabolites, 25-hydroxylamine-D[25(OH)D]
and 1$\alpha$, 25-(OH)2-D, occur with aging in both sexes\cite{dhaliwal2020effect,nordin1992osteoporosis}.
Nutritional vitamin D deficiency may contribute to secondary hyperparathyroidism and bone loss with
aging since decreases in serum 25(OH)D levels correlate inversely with serum parathyroid hormone levels
and positively with BMD\cite{khosla1997effects}.

As indicated in this study, hyperglycemia is another important factor determining BMD in patients with T2DM.
It could be explained as follows. Firstly, diabetes has been shown to cause decreased osteopsathyrosis, reduced bone formation, and enhanced osteoblast apophasis in a bone-loss mouse model\cite{he2004diabetes}. Secondly, hyperglycemia leads to glycosuria, which results in a loss of calcium. Hypercalciuria presented as a raised glomerular filtration rate, reduces calcium reabsorption  and impairs bone deposition in diabetic rats\cite{wang2019dapagliflozin}. Hypercalciuria decreases the level of calcium in the bone, leading to poor bone quality\cite{li2018positive}. Some reports indicate that the hypercalciuria in patients with uncontrolled blood glucose could stimulate parathyroid hormone secretion, which may contribute to the development of osteopenia\cite{riggs2002sex}. Thirdly, hyperglycemia is known to generate higher concentrations of AGEs in collagen\cite{yamagishi2005possible}. AGEs have been shown to be associated with decreased strength in human cadaver femurs\cite{bonds2006risk}. The combination of the accumulation of AGEs in bone collagen and lower bone turnover may contribute to reduced bone strength for a given BMD in diabetes\cite{schwartz2017efficacy}. AGEs and oxidative stress produced by hyperglycemia may reduce enzymatic beneficial cross-linking, inhibit osteoblast differentiation, and induce osteoblast apoptosis\cite{li2018positive}.

Height, weight, sex, and obesity are also factors affecting BMD in T2DM. As a Korean population-based study reported, sex affects BMD. The difference in BMD distribution at the same skeletal site may be partially explained by distinctive endocrine and paracone factors between the two sexes\cite{lei2004bone}. It has been suggested that bone loss in elderly men is mostly a result of decreased bone formation, whereas bone loss in postmenopausal women is a result of excessive bone resorption\cite{yan2002age}. Sex hormones may account for this difference. Estrogen rapidly decreases in postmenopausal women. An accelerated phase of predominantly cancellous bone loss initiated by menopause is the result of the loss of the direct restraining effect of estrogen on bone turnover\cite{cui2016assessment}. Estrogen acts on high-affinity estrogen receptors in osteoblast and osteoclasis to restrain bone turnover\cite{khosla1999osteoporosis}. Estrogen also regulates the production, by osteoblastic and marrow stromal cells, of cytokines involved in bone remodeling, such as interleukin (IL)-1, IL-6, tumor necrosis factor-$\alpha$, prostaglandin E2, transforming growth factor-$\beta$, eta, and osteoprotegerin\cite{khosla1999osteoporosis,jilka1998cytokines}. The net result of the loss of direct action of estrogen is a marked increase in bone resorption that is not accompanied by an adequate increase in bone formation, resulting in bone loss. The accelerated phase of bone loss in women is due to direct skeletal consequences of rapid reduction in serum estrogen following the menopause.%\cite{khosla1999osteoporosis}.

High body weight and obesity have been shown to be associated with high BMD in many observational studies\cite{shapses2012bone}. Obesity may lead to increased BMD because it is associated with higher 17 $\delta$ -estradiol levels and higher mechanical load, which may protect bone\cite{nelson2001estrogen,ohta1993differences}. Visceral fat accumulation is associated with higher levels of pro-inflammatory cytokinesis, which may up-regulate receptor activators of nuclear ligand, leading to increased bone resorption and therefore decreased BMD\cite{hofbauer2004clinical,smith2006systemic,campos2012role}.

Some studies have shown that weight loss, both intentional and unintentional, is associated with the decreases in BMD. The study by Geoffroy et al. has shown that more than 70\% of patients have clinically significant BMD loss at 12 months after bariatrics surgery\cite{geoffroy2019impact}. This loss of bone density was observed at the femoral neck and femur\cite{geoffroy2019impact}. Then the significant reduction in BMD was related by bivariate analysis to the extent of reduction in BMI, weight loss, and to loss of fat and lean mass\cite{geoffroy2019impact}. A recent study in elderly women has identified risk factors for hip BMD loss over four years and concluded that women who gain weight show attenuated BMD loss at the trochanter, femoral neck, and total hip\cite{gudmundsdottir2010risk}.

\section{Conclusion}
In this paper, we propose a new BN algorithm (PKCL) that can find the
underlying causal mechanism between six BDMs and their related
factors. PKCL includes two stages: the local stage that discovers
the local MB sets and the global stage that learns the direction of
casual-effect relationship.

In addition, to demonstrate the superiority and effectiveness of
PCKL, a clinical data set that concludes the clinical indexes of the
patient with T2DM was collected and preprocessed. Experiments on
this dataset shows that PKCL can discover the casual relationships
that have already been discovered in clinical literature. What's
more, PKCL can discover new casual relationships to assist clinical
researchers in carrying out new experiments, which can save a lot of
time and money. Different from other BN algorithm, PCKL incorporates
rich prior knowledge, which means it can achieve good performance
even when the dataset is small while the feature is numerous. What'
more, PCKL is not limited in the clinical literature but can be
adjusted into any domain if incorporated with prior knowledge. The
future work on this subject will employ other probabilistic models
\cite{chen2009probabilistic,chen2013efficient} and learning in the
model space \cite{chen2013learning,chen2013model} for this kind of
problems.

\begin{figure*}[htbp]
\centering
\includegraphics[height=8.0cm,width=15cm]{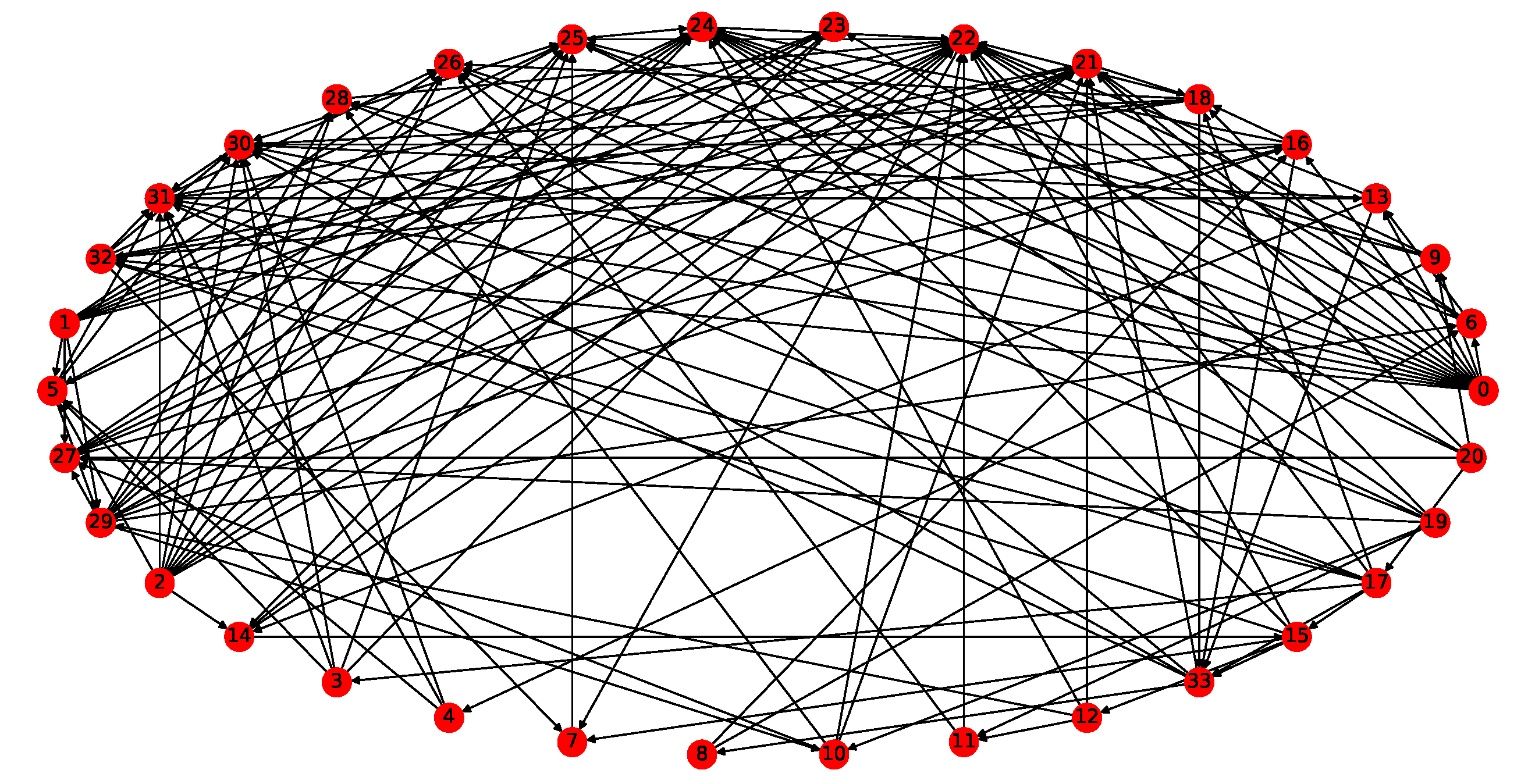}

\caption{The DAG before incorporating prior knowledge}
\end{figure*}

\begin{figure*}[htbp]
\centering
\includegraphics[height=8cm,width=15cm]{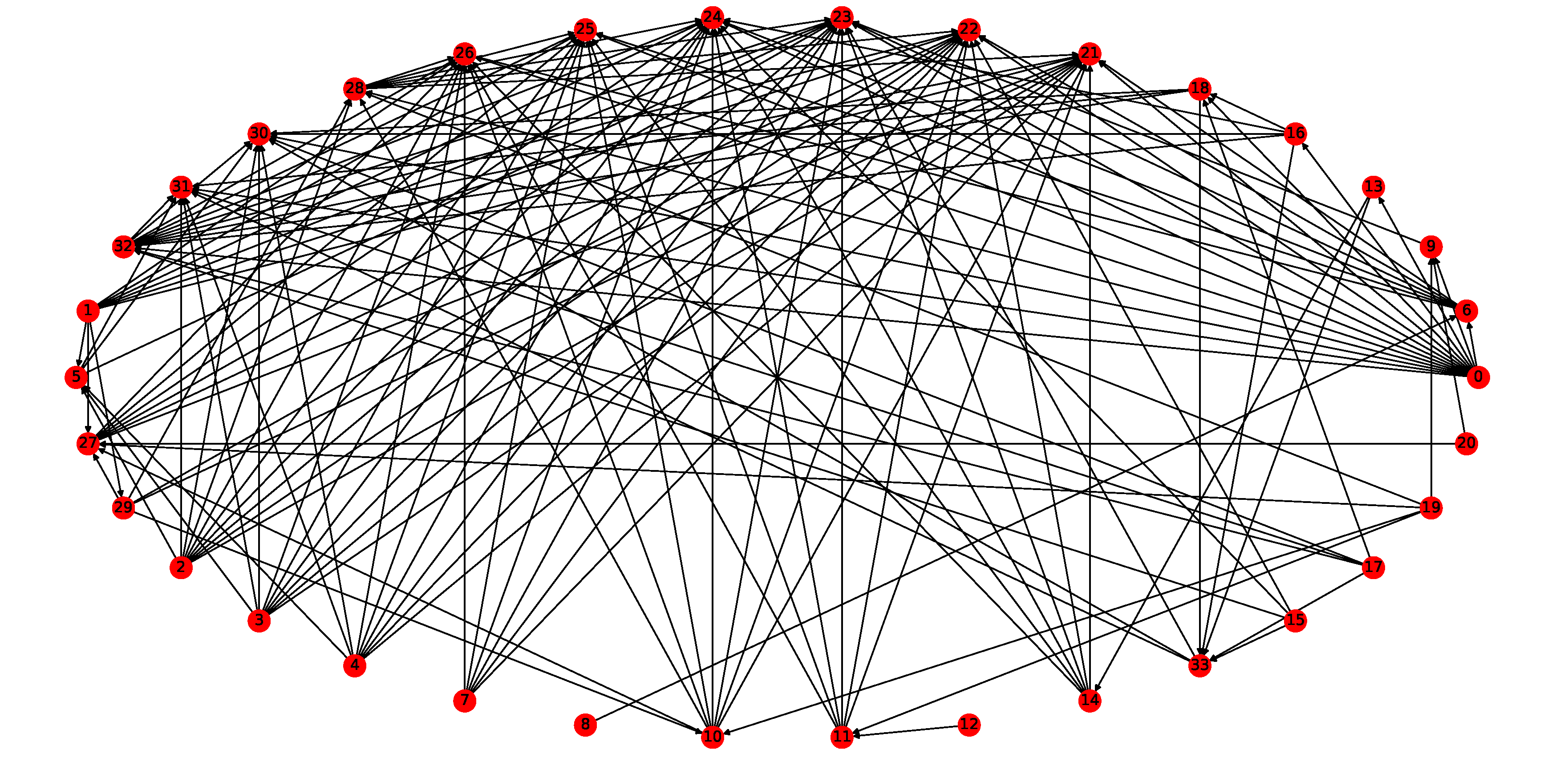}
\caption{The DAG after incorporating prior knowledge}
\end{figure*}

\section{Appendix}

\subsection{The process of discretization}

The discretization method that works as follows:
\begin{enumerate}
    \item Data is normalized so that each variable has mean 0 and standard deviation 1
    \item After normalization, association of each variable with the response variable is 
    computed using either Wilcoxon rank sum test (for binary response variable) or 
    Kruskal--Wallis non-parametric ANOVA (for multicategory response variable) at 0.05 alpha 
    level\cite{hollander2013nonparametric}.
    \item If a variable is not significantly associated with the response variable, it 
    is discretized as follows:
\begin{itemize}
    \item \emph{0 for values less than -1 standard deviation}
    \item \emph{1 for values between -1 and 1 standard deviation}
    \item \emph{2 for values greater than 1 standard deviation}
\end{itemize}
    \item If a variable is significantly associated with the response variable, it is discretized using
sliding threshold (into binary) or using sliding window (into ternary). The discretization threshold(s) is
determined by the Chi-squared test to maximize association with the response variable\cite{agresti2002categorical}.

\end{enumerate}

The discretization procedure can be instructed to compute necessary statistics only using training samples of the data to ensure unbiased estimation of error metrics on the testing data.

\subsection{Two pictures of the overall DAG}

Figure 2 is the learned DAG that not incorporating prior knowledge. Figure 3 is the learned DAG that incorporating prior knowledge. The circle represents the features and the arc represents the causal relationship. The numbers in the circle denote the features and Table 1 lists the corresponding relationships.

\subsection{Abbreviations and its descriptions}
Table 3 is the abbreviations appear in the paper and its descriptions 
\begin{table}[htbp]
\centering
\caption{Abbreviations and its descriptions}
 \begin{tabular}{lc}
  \toprule
Abbreviations & Descriptions \\
  \midrule
AGEs&advanced glycation end-products \\
BMD &bone mineral density \\
BN&Bayesian networks \\
CCMB&cross-check and complement MB discovery \\
CIs&conditional independent tests \\
DAG & Directed Acyclic Graph \\
DXA&X-ray absorptiometry \\
FS&feature selection \\
GS&The Growth and Shrink algorithm\\
IAMB&The increment associated MB algorithm \\
IPCMB&Iterative parent children-based MB \\
ITs&independent tests \\ 
kNN & k-Nearest Neighbors \\
KS& Koller-Sahami \\
MB&The Markov Blanket\\
MMMB&Min-max MB \\
NB & Naive Bayes\\
PC&parents and children \\
PCMB&Parent children-based MB algorithm  \\
PKCL & Prior-Knowledge-driven local Causal structure Learning \\
RF &Random Forest \\
SVM & Support Vector Machine \\
STMB&Simultaneous MB algorithm  \\
SL &structure learning  \\
T2DM &Type 2 diabetes\\
  \bottomrule
  \label{table3}
 \end{tabular}
\end{table}

\bibliographystyle{unsrt}
\bibliography{ref}

\end{document}